\documentclass[letterpaper]{article} 
\usepackage{aaai2026}  
\usepackage{times}  
\usepackage{helvet}  
\usepackage{courier}  
\usepackage[hyphens]{url}  
\usepackage{graphicx} 
\usepackage{natbib}  
\usepackage{caption} 
\usepackage{adjustbox}
\usepackage{float}
\usepackage{booktabs}
\usepackage{multirow}
\usepackage{tabularx}
\usepackage{float}
\usepackage{tikz}
\usepackage{pgfplots}\pgfplotsset{compat=1.18}
\usepgfplotslibrary{groupplots,fillbetween}
\usepackage{algorithm}
\usepackage{algpseudocode}
\usepackage{amsmath} 
\usepackage{subcaption}
\definecolor{TBlue}{HTML}{4E79A7}
\definecolor{TBlueLight}{HTML}{A0CBE8}
\definecolor{TOrange}{HTML}{F28E2B}
\definecolor{TGreen}{HTML}{59A14F}
\definecolor{TRed}{HTML}{E15759}
\definecolor{TPurple}{HTML}{B07AA1}
\definecolor{TGray}{HTML}{7F7F7F}
\newlength{\GroupSep}
\newlength{\SubplotW}
\usepackage{newfloat}
\usepackage{listings}
\DeclareCaptionStyle{ruled}{labelfont=normalfont,labelsep=colon,strut=off} 
\floatstyle{ruled}
\newfloat{listing}{tb}{lst}{}
\floatname{listing}{Listing}
\title{Reflection-Driven Control for Trustworthy Code Agents}
\author{
    Bin Wang\textsuperscript{\rm 1},
    Jiazheng Quan\textsuperscript{\rm 2},
    Xingrui Yu\textsuperscript{\rm 3,*},
    Hansen Hu\textsuperscript{\rm 1},
    Yuhao\textsuperscript{\rm 1},
    Ivor Tsang\textsuperscript{\rm 3}
}
\affiliations{
    \textsuperscript{\rm 1}School of Computer Science, Peking University, China \\
    \textsuperscript{\rm 2}Xiamen University, China \\
    \textsuperscript{\rm 3}Centre for Frontier AI Research (CFAR), Agency for Science, Technology and Research (A*STAR) \\
    thebinking66@stu.pku.edu.cn, 
    jiazhengquan@stu.xmu.edu.cn,
    yu\_xingrui@cfar.a-star.edu.sg,
    hansen\_sun@stu.pku.edu.cn,
    hyu25@stu.pku.edu.cn,
    ivor\_tsang@cfar.a-star.edu.sg \\
        \textsuperscript{*}Corresponding author.

}

\usepackage{bibentry}

\begin{document}

\maketitle

\begin{abstract}

Contemporary large language model (LLM) agents are remarkably capable, but they still lack reliable safety controls and can produce unconstrained, unpredictable, and even actively harmful outputs. To address this, we introduce \textbf{Reflection-Driven Control}, a standardized and pluggable control module that can be seamlessly integrated into general agent architectures. Reflection-Driven Control elevates “self-reflection” from a post hoc patch into an explicit step in the agent’s own reasoning process: during generation, the agent continuously runs an internal reflection loop that monitors and evaluates its own decision path. When potential risks are detected, the system retrieves relevant repair examples and secure coding guidelines from an evolving reflective memory, injecting these evidence-based constraints directly into subsequent reasoning steps. We instantiate Reflection-Driven Control in the setting of secure code generation and systematically evaluate it across eight classes of security-critical programming tasks. Empirical results show that Reflection-Driven Control substantially improves the security and policy compliance of generated code while largely preserving functional correctness, with minimal runtime and token overhead. Taken together, these findings indicate that Reflection-Driven Control is a practical path toward trustworthy AI coding agents: it enables designs that are simultaneously autonomous, safer by construction, and auditable.

\end{abstract}

\section{Introduction}

Autonomous LLM agents are rapidly evolving from single-turn text generators into systems capable of multi-step task execution, tool use, and real-world decision making \cite{chowa2025language}. This shift fuses internal reasoning with external actions, enabling agents to browse at scale, invoke APIs, and modify external artifacts, thereby exhibiting cross-task competence. As a result, the design space is expanding quickly, while challenges around safety, control, and evaluation are intensifying. With long-term memory, proactive planning, and environment interaction becoming core components, AI agents are transforming from single-model reasoners into distributed networks of cooperating components. This transition is reshaping generative AI practice and catalyzing a global discussion on agentic AI security \cite{park2023generative}.

Despite this progress, contemporary agent workflows still produce untrusted content in uncontrollable ways. Even strong base models can hallucinate or emit unsafe outputs. When tools and autonomy enter the loop, jailbreaks, prompt injection, and agent worms further expose fragile control surfaces \cite{john2025owasp}. Such failures can translate directly into system risk and unsafe behavior. 

Advancing the safety and trustworthiness of agent workflows hinges on two core challenges. \textbf{(i) Trusted control at decision time}: how to dynamically constrain agent behavior during reasoning and execution to prevent task drift and hazardous tool calls. \textbf{(ii) Post-hoc verifiability and auditability}: how to trace the evidential basis and execution logic of decisions, thereby enabling accountability and transparency.

We propose a standardized reflect agent module that elevates “reflection” from an external, post-hoc procedure to an internal, first-class control loop within the agent. By tightly coupling the reflection pathway with the knowledge feedback channel, the agent architecture equipped with this module can perform continuous self-supervision and self-correction across the planning, execution, and verification stages, thereby significantly enhancing system trustworthiness and auditability without compromising autonomy.

To concretely validate the system-level benefits of reflection-centric orchestration, we instantiate our framework in \emph{secure code generation}. Our experiments show that, relative to strong agent baselines, embedding a reflection agent yields consistent gains in \emph{safety}, \emph{consistency}, and \emph{traceability}, resulting in higher practical trust and auditability. This evidences that the reflection-centric control is effective and scalable for high-risk tasks.

The contributions of this article mainly include the following three points:

\begin{itemize}
\item \textbf{A reflection-driven, closed-loop control framework.} We integrate reflection as a first-class control circuit that spans planning, execution, and verification, with an auditable evidence trail rather than ad-hoc post-processing, and implemente it as the Reflect layer in a Plan–Reflect–Verify agentic framework.
\item \textbf{A practical instantiation for secure code generation.} We compose lightweight self-checks, dynamic memory/RAG, reflective prompting, and tool governance (compiler/tests/CodeQL) into an evidence-grounded generation pipeline that maintains autonomy while enforcing safety.
\item \textbf{Comprehensive evaluation and analysis.} On public security-oriented code-generation benchmarks with strict compile/run/static-analysis validation, our framework delivers consistent improvements over agent baselines, alongside ablations and case studies that quantify each component’s impact and illuminate failure modes in high-risk settings.
\end{itemize}

\section{Related Work}

\paragraph{Agent framework and system security}

As LLMs evolve from passive generators to autonomous decision-makers, agentic systems have become central to generative AI.
\citet{han2024llm} note that multi-agent systems enable complex task coordination through role division and communication but still lack robust safety boundaries. Thus, while multi-agent architectures enhance automation and scalability in security assessment, they can also introduce systemic risks, where coordination failures or misaligned goals may propagate across agents \cite{david2025multi}. Ensuring cross-agent safety constraints, auditable decisions, and real-time intervention is therefore essential.

Recent work shows that the attack surface of agentic systems exceeds that of traditional AI, encompassing prompt manipulation, toolchain hijacking, external API abuse, and persistent session pollution \cite{achiam2023gpt}.
The THOR framework \cite{narajala2025securing} systematizes these threats across the agent lifecycle and argues for layered defenses and auditability to ensure traceable safety. More broadly, \citet{de2025open} formalize multi-agent security, showing that decentralized cooperation among autonomous agents introduces systemic risks such as covert collusion, coordinated group attacks, and cross-platform propagation. These observations suggest a shift in security governance from single-model safeguards to interaction-level defenses among agents.

The RepairAgent \cite{bouzenia2024repairagent} exemplifies safe autonomy by invoking compilation, testing, and verification tools under strict system constraints, demonstrating that tool-integrated architectures can maintain auditability.
Overall, research on intelligent agents is shifting from task accuracy to safety-by-design. The key challenge lies in embedding intrinsic safety constraints and self-supervision, the foundation of trustworthy Agentic AI.

\paragraph{TRiSM for agentic AI}

The autonomy and interactivity of agentic systems improve flexibility but broaden the attack surface, introducing new security risks.
These systems thus require evaluation under the Trust, Risk, and Security Management (TRiSM) framework \cite{raza2025trism}.
Recent studies reveal vulnerabilities to jailbreaks, prompt injections, and tool misuse that can cascade through multi-agent networks \cite{john2025owasp,narajala2025securing}.
\citet{huang2023catastrophic} show that even safety-aligned models can suffer up to 95\% misalignment via generative exploitation, while the DeepInception exposes cognitive vulnerabilities through contextual prompts \cite{li2023deepinception}.
TRiSM further warns of “prompt infection” attacks, where malicious instructions propagate among agents \cite{lee2024prompt}.

To mitigate these threats, research has moved from reactive defense to design-level resilience.
In the input layer, the self-reminder mechanism reduced jailbreak rates from 67.21\% to 19.34\% by injecting safety cues \cite{xie2023defending}, while the decoupled refusal training enables mid-generational rejection of unsafe outputs \cite{yuan2024refuse}.
At the system level, \citet{debenedetti2025defeating} proposed CaMeL to enforce the least privilege via role–tool binding and flow isolation, complemented by hierarchical monitoring \cite{wang2025megaagent} for behavior tracking and auditability.

Despite these progresses, the interpretability and privacy of TRiSM remain weak.
Layered-CoT \cite{sanwal2025layered} improves the transparency of reasoning but lacks real-time threat detection, while differential privacy and federated learning do not ensure complete inter-agent isolation.
In general, current defenses provide fragmented protection across input and output layers, underscoring the need for a unified, closed-loop security control that integrates reasoning, planning, and tool execution for trustworthy Agentic AI.

\paragraph{Trust and verifiability in agentic systems}

Trustworthiness and verifiability are central to agent safety governance. Narajala et al.’s THOR framework shows that generative-agent risks span the entire lifecycle, requiring layered auditing and traceable trust chains \cite{narajala2025securing}.
Expanding on this, RAS-Eval \cite{fu2025ras} introduces a real-world benchmark with 11 CWE categories to test LLM agents in dynamic tasks, where attacks reduced success rates by 36.78\%, exposing weak task-control mechanisms. CyberSOCEval \cite{deason2025cybersoceval}, targeting malware analysis and threat reasoning, found that LLMs have reasoning strength but lack transferable defensive capabilities, demanding task-specific optimization.

The DeepSeek-R1 model \cite{guo2025deepseek} achieved closed-source–level reasoning, suggesting reinforced reasoning can support verifiable agents though excessive autonomy adds risk.
In summary, trust verification is becoming a core requirement for safe Agentic AI. From THOR’s layered defense to RAS-Eval and CyberSOCEval’s task-level evaluations, future frameworks must ensure interpretable reasoning, auditable execution, and verifiable reflection to balance autonomy with control.

\section{Problem Formulation and Design Goals}

To systematically evaluate the trustworthiness of code generation agents, we first construct a standardized, plug-and-play module that can be integrated with general agentic orchestration, and use it to build a baseline code generation agent architecture as our evaluation benchmark. As illustrated in Figure~\ref{fig:reflex_overview}(a), the left side shows the standardized module, while the right side presents the baseline agent architecture.

Building upon this foundation, this work focuses on enhancing the safety and trustworthiness of the code generation process. We formulate the core problem as follows: given a code snippet \( x \) with potential security flaws and its contextual environment (including file-level context \( C_f \) and function-level context \( C_{fn} \)), the objective is to generate a repaired code snippet \( y \) that must satisfy:
\begin{enumerate}
    \item \textbf{Security}: Eliminate common security risks present in \( x \) (e.g., SQL injection, command injection, buffer overflow, path traversal, XSS, null pointer dereference, integer overflow, and use-after-free).
    \item \textbf{Functionality}: Preserve the semantic intent of the original code \( x \).
    \item \textbf{Executability}: Ensure the generated code \( y \) is compilable or interpretable.
\end{enumerate}

We formalize this secure code generation task as a conditional generation process:
\begin{equation}
\label{eq:problem_formulation}
y = \mathcal{G}\big( x, C_f, C_{fn}, \mathcal{R}(x, C_f, C_{fn}; \mathcal{M}), \theta \big),
\end{equation}
where \( \mathcal{G} \) denotes the large language model generator, \( \mathcal{R} \) is a retrieval function that fetches high-value security repair exemplars from a memory repository \( \mathcal{M} \) as reflective evidence, and \( \theta \) represents the model and prompting parameters.

The overarching goal of this research is to achieve autonomous and trustworthy LLM-based code generation. The core mechanism involves an agent proactively proposing code edit patches that must be \textbf{secure}, \textbf{policy-compliant}, and \textbf{auditable}. Diverging from approaches reliant on explicit planning, we introduce a Reflex Module to realize two critical functions: (i) serving as a security gate to intercept unsafe code changes before final commitment, and (ii) generating machine-verifiable evidence to endow the agent's decision-making and repair actions with auditability.

Based on the aforementioned problem formulation, the design goals of this work are concretely summarized as follows:
\begin{itemize}
    \item[(\textbf{G1})] \textbf{High Security Efficacy}: Minimize the vulnerability rate and policy violation rate of the generated code.
    \item[(\textbf{G2})] \textbf{Verifiability}: Provide structured, machine-verifiable audit evidence for each repair operation.
    \item[(\textbf{G3})] \textbf{Lightweight Integration}: Maintain a design free from complex planning, ensuring the Reflex Module can be seamlessly integrated into existing coding agent architectures in a plug-and-play manner with low overhead.
\end{itemize}

\section{Reflex Module: Architecture and Implementation}
\begin{figure*}[t]
    \centering
    \resizebox{0.95\textwidth}{!}{%
    \begin{minipage}{\textwidth}
        \begin{subfigure}[t]{0.47\textwidth}
            \centering
            \includegraphics[width=\linewidth]{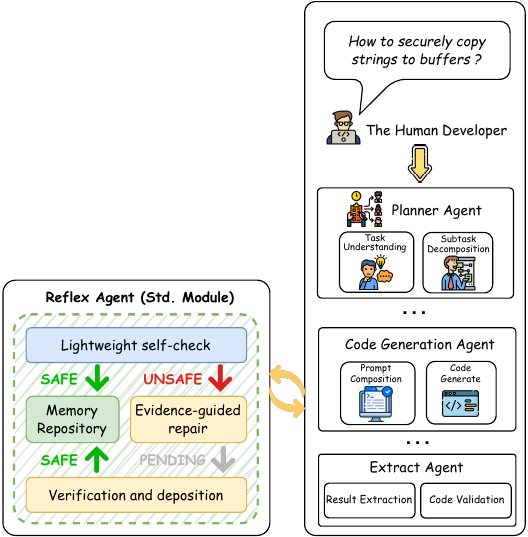}
            \caption{Reflex Agent as a plug-and-play module in general agent orchestration.}
            \label{fig:reflex_integration}
        \end{subfigure}
        \hfill
        \begin{subfigure}[t]{0.52\textwidth}
            \centering
            \includegraphics[width=\linewidth]{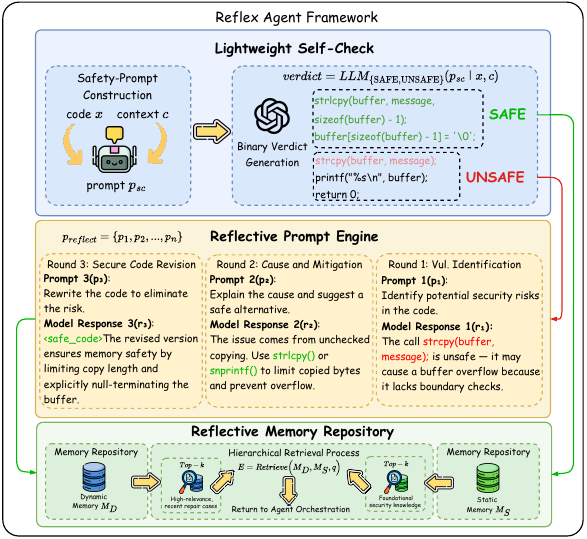}
            \caption{Overall framework of the Reflex Agent.}
            \label{fig:reflex_framework}
        \end{subfigure}
    \end{minipage}%
    }
    \caption{(a) shows the plug-and-play integration of the Reflex Agent within a general multi-agent orchestration.
    (b) presents the internal framework of the Reflex Agent, illustrating its modular design and workflow.}
    \label{fig:reflex_overview}
\end{figure*}

We instantiate the \textsc{Reflex} module as the Reflect layer of a agentic framework, where reflection functions as a first-class control circuit supervising planning, execution, and verification. It is integrated into the end-to-end code generation pipeline as a modular, pluggable component (equivalently, an agent submodule). The core workflow is:
\begin{enumerate}
\item \textbf{Lightweight self-check.} A fast pre-checker screens the input code to avoid unnecessary computation.
\item \textbf{Evidence-guided repair.} If the input is judged \emph{unsafe}, the system retrieves high-value prior cases from a dynamic memory and fuses them with best-practice constraints into a reflection prompt that steers the model to generate an evidence-based fix.
\item \textbf{Verification and deposition.} The candidate output is filtered and compilation-checked; verified artifacts are then written back to memory, forming a continuously evolving knowledge loop.
\end{enumerate}
This design realizes a closed loop of \emph{low-cost, frontloaded reflection}, \emph{evidence-driven generation}, and \emph{auditable knowledge accumulation}, without fine-tuning the underlying models.
Building on this design, we implement a standard \textsc{Reflex} agent (see Figure~\ref{fig:reflex_overview}) that plugs into the system with three core components: (i) a \emph{Lightweight Self-Checker}, (ii) a \emph{Reflective Prompt Engine}, and (iii) a \emph{Reflective Memory Repository}. We detail each component below.

\subsection{Lightweight Self-Check and Routing Mechanism}

The lightweight self-checker serves as a front-end filtering layer for the Reflex module, responsible for performing an initial, trustworthy safety diagnosis of the input code. This component is designed to evaluate the potential safety status of a code snippet at extremely low computational cost, thereby enabling an efficient task-routing mechanism.

Specifically, we model the self-check process as a binary classification task based on a large language model. Given the input code $x$ and its contextual information $c$, the system constructs a concise safety-review prompt $p_{\mathrm{sc}}$, instructing the model to output only a binary judgment: “SAFE” or ``UNSAFE''. Formally, this process can be represented as:

\begin{equation}
\text{verdict} = \mathrm{LLM}_{\{\text{SAFE},\, \text{UNSAFE}\}} \left( p_{\mathrm{sc}} \mid x, c \right).
\label{eq:verdict}
\end{equation}

If the verdict is determined as ``SAFE'', the sample and its metadata are directly written into the dynamic memory repository as a retained safe case. Conversely, if the verdict is ``UNSAFE'', the system triggers the full reflection and repair pipeline, where the reflection prompt engine (introduced later) performs in-depth code diagnosis and correction.

This mechanism effectively routes trustworthy responses without redundant reflection processes, thereby reducing the average inference cost of the system. It is particularly advantageous for large-scale task processing scenarios, where efficient filtering and resource allocation are critical.

\subsection{Reflective Prompt Engine}

The Reflective Prompt Engine serves as the core driving module of the reflection process, enabling the agent to perform deep analysis and self-improvement on problematic code. When a code segment is classified as ``UNSAFE'' and enters the reflection phase, the engine constructs a structured, multi-turn prompt, transforming a single code generation task into a multi-round chain-of-thought reflection dialogue.

During this process, the model is guided to conduct systematic introspection and reasoning on the problematic code and its context, identifying vulnerabilities and devising corresponding repair strategies. This multi-turn reflective dialogue ensures that the complete reasoning chain—from problem recognition to solution implementation—is explicitly unfolded and recorded, thereby achieving knowledge distillation for code generation tasks.

Under the operation of the Reflective Prompt Engine, the resulting structured reflection records are systematically stored in the dynamic memory repository. In essence, each resolved case is elevated into a reusable reasoning pattern, offering high-quality reference for future related issues. As more solutions and reflective processes accumulate, the system’s knowledge base evolves continuously, enhancing both the quality of stored knowledge and the agent’s adaptive capability in subsequent tasks. This growing body of experience strengthens the overall trustworthiness and transparency of the intelligent agent framework.

\subsection{Reflective Memory Repository}

The Reflective Memory Repository is the central component enabling the system’s continual learning and experience reuse. By integrating vectorized retrieval with structured metadata management, it constructs a dynamically evolving knowledge base for safety repair and reflection.

To balance retrieval efficiency with comprehensive knowledge coverage, we design a hierarchical retrieval architecture composed of two layers:

\begin{itemize}

    \item \textbf{Dynamic Memory \(M_D\)}: Implemented with a ChromaDB vector database to store and index verified repair cases produced during system operation. Its strengths are high relevance and low retrieval latency.
    \item \textbf{Static Memory \(M_S\)}: Built from predefined secure coding standards, vulnerability databases, and other static knowledge artifacts, serving as the system’s foundational knowledge anchor to ensure completeness of core principles.
\end{itemize}

Given a query \(q\) composed of the current problematic code \(x\) and its context \(c\), the hierarchical retrieval process 
Retrieve \(\text{Retrieve}(M_D, M_S, q)\) is formally defined as follows:

\begin{equation}
\label{eq:retrieve}
\begin{aligned}
E & = \text{Retrieve}(M_D, M_S, q) \\ & = 
\begin{cases}
\text{Top-}k\ \text{from } M_D, & \text{if } condition \\

\text{Top-}k\ \text{from } M_D \cup M_S, & \text{otherwise}
\end{cases} 
\end{aligned}
\end{equation}
Here, \(E\) denotes the final set of retrieved experiences, \(|E_{M_D}|\) represents the number of results recalled from \(M_D\), \(\text{sim}(\cdot)\) denotes the cosine similarity function, \(k_{\text{min}}\) is the minimum recall threshold, and \(\theta\) is the lower bound for similarity. The $condition$ is defined as $|E_{M_D}| \geq k_{\text{min}} \ \text{and} \ \max(\text{sim}(E_{M_D}, q)) \geq \theta$.

This retrieval strategy prioritizes high-value, high-relevance, and recent experiences from \(M_D\), ensuring that the system can rapidly reuse validated contextual knowledge. When the dynamic memory fails to meet retrieval requirements, such as insufficient results or low similarity scores, the system automatically falls back to the static memory \(M_S\) for supplementary queries, thus maintaining both adaptivity and knowledge completeness in the overall reflection framework.

\section{Experiment}

To comprehensively evaluate the effectiveness and efficiency of the proposed reflection module, we formulate the following three Research Questions (RQs). These questions are designed to probe the system from three complementary dimensions: performance gain, mechanistic depth, and computational overhead.

\begin{description}

\item[\textbf{RQ1: Effectiveness Improvement Analysis}] 
\emph{To what extent does the introduction of the reflection module enhance the base agent’s overall trustworthiness in terms of code safety and functional correctness?}

This question aims to quantify the core performance gains brought by the reflection module. We empirically compare the base agent’s performance under two configurations, with and without the reflection module, focusing on key indicators such as Security Rate and Pass Rate. The objective is to assess the module’s impact on the overall quality of code generation.

\item[\textbf{RQ2: Reflection Depth and Diminishing Returns}] 
\emph{Is performance improvement positively correlated with the depth (or number of rounds) of reflection? Does a point of diminishing returns exist?}

This question investigates the operational boundary and marginal utility of the reflection mechanism. By controlling the maximum number of reflection rounds and tracking the evolution of performance metrics (e.g., Security Rate) after each iteration, we aim to identify the optimal reflection depth, beyond which further computational investment yields negligible improvement.

\item[\textbf{RQ3: Cost–Benefit Analysis}]
\emph{What are the additional computational and temporal costs associated with the performance gains introduced by the reflection module?}

This question evaluates the practical efficiency of the system. By measuring and comparing the total token consumption (reflecting API cost) and total task execution time (reflecting latency) between the two configurations, with and without reflection, we precisely quantify the trade-off between performance enhancement and resource expenditure.

\end{description}

\subsection{Benchmark}

To evaluate the improvements brought by the Reflex Agent Module in code generation safety and reliability, we selected a set of challenging scenarios from the most influential Common Weakness Enumeration (CWE) categories. We adopted the dataset validated by \cite{he2023large}, which covers eight representative scenarios drawn from MITRE’s list of the Top 25 Most Dangerous Software Weaknesses. The specific scenarios can be found in the Appendix Table~\ref{tab:cwe_scenarios}. 

Each CWE scenario includes two to three carefully designed programming environments, refined by He et al. to eliminate low-quality prompts and to emulate a variety of real-world code completion tasks. This makes the dataset a powerful benchmark for assessing a model’s ability to generate secure and trustworthy code. The tasks involve incomplete C/C++ or Python code prompts, challenging the model to produce appropriate completions that demonstrate its capacity to handle partial or ambiguous inputs in practical programming contexts.
For every model, 25 code samples are generated per scenario and repeated across five independent runs to minimize experimental variance and ensure objective evaluation.

All generated code samples are analyzed using the static analysis tool CodeQL \cite{szabo2023incrementalizing} to detect vulnerabilities and assess the impact of generated repairs. In addition, we employ a trustworthy safety compliance evaluation system based on LLM judges to comprehensively assess code quality, security, and compliance, which complements traditional SAST tools by covering aspects of trustworthiness that static analyzers cannot fully capture.

\begin{table*}[!h]
\centering
\caption{Reflex module agent performance summary.}
\label{tab:Reflextor_summary}
\renewcommand{\arraystretch}{1.15}
\resizebox{1.4\columnwidth}{!}{ 
\begin{tabular}{lcccccccc}
\toprule
\multirow{2}{*}{\textbf{Metric}} & \multicolumn{2}{c}{\textbf{gpt-3.5-turbo}} & \multicolumn{2}{c}{\textbf{gpt-4o}} & \multicolumn{2}{c}{\textbf{qwen3-coder-plus}} & \multicolumn{2}{c}{\textbf{gemini-2.5-pro}} \\
\cmidrule(r){2-3} \cmidrule(r){4-5} \cmidrule(r){6-7} \cmidrule(r){8-9}
& Base & Base+Reflex & Base & Base+Reflex & Base & Base+Reflex & Base & Base+Reflex \\
\midrule
Sec. Rate     & 93.7 & 96.6 (\textbf{$\uparrow$2.9})  & 85.7 & 95.0 (\textbf{$\uparrow$9.3})    & 83.7 & 94.9 (\textbf{$\uparrow$11.2})  & 88.0 & 97.1 (\textbf{$\uparrow$9.1}) \\
Pass Rate     & 88.0 & 92.4 (\textbf{$\uparrow$4.4})  & 95.2 & 94.9 ($\downarrow$0.3)           & 86.7 & 80.1 ($\downarrow$6.6)          & 91.4 & 94.9 (\textbf{$\uparrow$3.5}) \\
Eff. Total    & 22.0 & 23.1 (\textbf{$\uparrow$1.1})  & 23.8 & 23.7 ($\downarrow$0.1)           & 21.6 & 20.2 ($\downarrow$1.4)          & 22.9 & 23.7 (\textbf{$\uparrow$0.8}) \\
Sec. Count    & 23.4 & 24.1 (\textbf{$\uparrow$0.7})  & 21.4 & 23.8 (\textbf{$\uparrow$2.4})    & 17.9 & 23.7 (\textbf{$\uparrow$5.8})   & 22.0 & 24.3 (\textbf{$\uparrow$2.3}) \\
Unres. Count  & 3.0  & 1.9  ($\downarrow$1.1)         & 1.2  & 1.3  (\textbf{$\uparrow$0.1})    & 3.3  & 4.8  (\textbf{$\uparrow$1.5})   & 2.1  & 1.3  ($\downarrow$0.8) \\
\bottomrule
\end{tabular}
}
\end{table*}

\subsection{Metrics}

To comprehensively evaluate the proposed method across the three key dimensions of code safety, functionality, and practicality, we define a unified evaluation framework that integrates both quantitative and qualitative metrics.

Appendix Table~\ref{tab:quant-metrics} and Table~\ref{tab:qual-metrics} define the full evaluation protocol used in this work. Table~\ref{tab:quant-metrics} reports the quantitative metrics used for all models, capturing (i) whether the model can generate code that actually builds, (ii) whether that code is functionally correct, and (iii) whether that code is secure under static analysis. All percentage-based metrics are computed only over compilable samples, so they reflect realistic, runnable code rather than idealized snippets. Table~\ref{tab:qual-metrics} complements this with qualitative dimensions of production readiness: overall engineering quality (readability, modularity, maintainability), defensive completeness against common exploit patterns, and policy / privacy compliance. Each compiled sample is reviewed along these axes, allowing us to surface risks that automated scanners such as CodeQL may not fully capture.

\subsection{RQ1: Effectiveness Improvement Analysis}

To verify the effectiveness of the Reflex Module Agent in enhancing the code safety and functional reliability of the base agents, we conducted a systematic evaluation across four mainstream LLMs. The experiments covered eight key CWE vulnerability categories, each containing 2–3 specific programming scenarios. For each model and scenario, 25 code samples were generated and tested over five independent experimental runs to ensure statistical significance of the results.

Integrating the Reflex Module Agent leads to substantial overall improvements in both safety and functionality of the generated code.
As shown in Table \ref{tab:Reflextor_summary}, the agent architectures augmented with the Reflex Module consistently achieved significant gains in Security Rate (Sec.Rate) compared to their respective base versions across all tested models. 

These results indicate that the proposed framework effectively enhances code safety generation capabilities across diverse LLM architectures and scales, reinforcing its generalizability and robustness in secure code synthesis.

\paragraph{The dynamic memory repository within the reflection mechanism serves as the core driver of performance improvement.}
To further investigate the source of these gains, we closely analyzed the evolution of the dynamic RAG subsystem. As shown in Table \ref{tab:dynamic_rag}, across five consecutive experimental rounds, the dynamic RAG system exhibited clear signs of knowledge accumulation and convergence.

\begin{table}[h]
\footnotesize
\centering
\caption{
Evolution of retrieval performance during dynamic RAG iterations in the framework. 
\textit{ARD} is the average number of retrieved documents; 
\textit{ASim}, \textit{MSim}, and \textit{mSim} represent the average, maximum, and minimum cosine similarities, respectively; 
\textit{RSR} is the retrieval success rate; 
and \textit{FUR} denotes the fallback usage rate triggered when retrieval confidence is insufficient.
}
\label{tab:dynamic_rag}
\begin{tabular}{lcccccc}
\toprule
\textbf{Iteration} & \textbf{ARD} & \textbf{ASim} & \textbf{MSim} & \textbf{mSim} & \textbf{RSR} & \textbf{FUR} \\
\midrule
Run 1 & 2.1 & 0.850 & 1.000 & 0.715 & 85\% & 15\% \\
Run 2 & 2.8 & 0.950 & 1.000 & 0.716 & 95\% & 5\% \\
Run 3 & 2.9 & 0.960 & 1.000 & 0.716 & 98\% & 2\% \\
Run 4 & 3.0 & 0.970 & 1.000 & 0.746 & 100\% & 0\% \\
Run 5 & 2.9 & 0.980 & 1.000 & 0.716 & 100\% & 0\% \\
\bottomrule
\end{tabular}
\end{table}

The average retrieval similarity steadily increased from 0.850 in the first round to 0.980 in the fifth round, marking a 15.3\% improvement. Simultaneously, the system’s autonomy strengthened notably: the retrieval success rate rose from 85\% to 100\%, while the fallback rate to the static memory decreased from 15\% to 0\%.

These results indicate that through experience reuse, the system achieved a stable and self-sustaining state around the fourth round, effectively forming an autonomous, continuously evolving safety knowledge base. This self-reinforcing dynamic memory mechanism is thus the key enabler of sustained efficiency and safety enhancement in the Reflex framework.

\paragraph{Retrieval quality is strongly correlated with repair effectiveness.}

We further examined the relationship between document similarity in retrieval and the accuracy of security repairs (see Table \ref{tab:saferagen}). The data reveal a clear positive correlation: when the retrieved document’s similarity exceeded the 0.70 threshold, the corresponding repair accuracy reached over 93.8\%. In contrast, for documents with similarity below 0.70, the accuracy dropped to 75.0\%.
Notably, within the high-similarity range (0.95–1.00), the repair accuracy achieved 100\%, indicating perfect alignment between retrieved experiences and the current task.

\begin{table}[!h]
\footnotesize
\centering
\caption{
Distribution of retrieved document quality across different similarity intervals in the framework. 
\textit{Sim. Range} denotes the cosine similarity interval; 
\textit{Docs} is the number of retrieved documents within each range; 
\textit{Ratio} indicates their proportion in all retrieved samples; 
and \textit{FixAcc} represents the accuracy of security patch generation associated with documents in that range.
}
\label{tab:saferagen}
\begin{tabular}{lccc}
\toprule
\textbf{Sim. Range} & \textbf{Docs} & \textbf{Ratio} & \textbf{FixAcc} \\
\midrule
0.95--1.00 & 187 & 62.3\% & 100\% \\
0.85--0.95 & 73 & 24.3\% & 98.6\% \\
0.70--0.85 & 32 & 10.7\% & 93.8\% \\
$<$0.70 & 8 & 2.7\% & 75.0\% \\
\bottomrule
\end{tabular}
\end{table}

These findings validate the effectiveness of the similarity threshold (0.70) used in our system. It strikes a well-calibrated balance between recall and precision, ensuring that retrieved experiences remain both relevant and reliable, thereby sustaining high-quality knowledge reuse within the dynamic RAG framework.

\subsection{RQ2: Reflection Depth and Diminishing Returns}
To explore the relationship between the depth of reflection (i.e., number of iterative rounds) and the corresponding performance gains, as well as to identify potential points of diminishing returns, we conducted multi-round iterative experiments and continuously monitored key performance indicators.

The reflection mechanism exhibits a clear convergence point of diminishing returns.
Analysis of the dynamic RAG subsystem (see Appendix Table \ref{tab:evolution}) shows that retrieval performance stabilized after the fourth iteration. The average similarity increased from 0.850 to 0.980, after which further gains were negligible, and the fallback rate dropped to zero. This indicates that once approximately 100 high-quality samples (4 rounds × 25 scenarios) were accumulated, the dynamic memory repository reached a state of knowledge saturation, effectively covering the majority of retrieval demands.

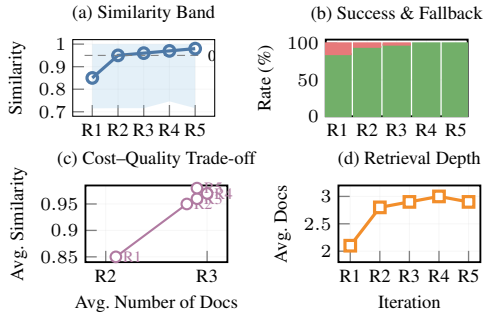
\begin{figure}[!h]
\centering
\scriptsize

\begin{tikzpicture}[
    /pgfplots/every axis/.style={
        line width=0.7pt,
        tick style={line width=0.5pt},
        tick label style={font=\scriptsize},
        label style={font=\scriptsize},
        title style={font=\scriptsize, yshift=-2pt}
    }
]
\begin{groupplot}[
    group style={
        group size=2 by 1,
        horizontal sep=\GroupSep,
        vertical sep=6pt
    },
    width=\SubplotW,
    height=0.75\SubplotW,
    tick pos=left,
    grid=both,
    grid style={opacity=0.12},
    xtick={1,2,3,4,5},
    xticklabels={R1,R2,R3,R4,R5}
]

\nextgroupplot[
    ylabel={Similarity},
    ymin=0.68, ymax=1.02,
    title={(a) Similarity Band}
]

\addplot[name path=minsim, draw=none] coordinates {
    (1,0.715) (2,0.716) (3,0.716) (4,0.746) (5,0.716)
};
\addplot[name path=maxsim, draw=none] coordinates {
    (1,1.000) (2,1.000) (3,1.000) (4,1.000) (5,1.000)
};
\addplot[fill=TBlueLight, fill opacity=0.28, draw=none]
    fill between[of=minsim and maxsim];

\addplot[very thick, TBlue, mark=o, mark options={fill=white}] coordinates {
    (1,0.850) (2,0.950) (3,0.960) (4,0.970) (5,0.980)
};

\addplot[TGray, densely dashed] coordinates {(0.7,0.95) (5.3,0.95)};
\node[anchor=west, font=\tiny] at (axis cs:5.25,0.95) {0.95};

\nextgroupplot[
    ylabel={Rate (\%)},
    ymin=0, ymax=105,
    ymajorgrids=true,
    title={(b) Success \& Fallback}
]
\addplot[ybar stacked, fill=TGreen!85, draw=TGreen!85] coordinates {
    (1,85) (2,95) (3,98) (4,100) (5,100)
};
\addplot[ybar stacked, fill=TRed!80, draw=TRed!80] coordinates {
    (1,15) (2,5) (3,2) (4,0) (5,0)
};

\end{groupplot}
\end{tikzpicture}

\begin{tikzpicture}[
    /pgfplots/every axis/.style={
        line width=0.7pt,
        tick style={line width=0.5pt},
        tick label style={font=\scriptsize},
        label style={font=\scriptsize},
        title style={font=\scriptsize, yshift=-2pt}
    }
]
\begin{groupplot}[
    group style={
        group size=2 by 1,
        horizontal sep=\GroupSep,
        vertical sep=6pt
    },
    width=\SubplotW,
    height=0.75\SubplotW,
    tick pos=left,
    grid=both,
    grid style={opacity=0.12},
    xtick={1,2,3,4,5},
    xticklabels={R1,R2,R3,R4,R5}
]

\nextgroupplot[
    xlabel={Avg. Number of Docs},
    ylabel={Avg. Similarity},
    xmin=1.8, xmax=3.2,
    ymin=0.84, ymax=0.985,
    title={(c) Cost–Quality Trade-off}
]
\addplot[
    thick, TPurple, mark=*,
    mark options={fill=white},
    nodes near coords,
    every node near coord/.style={font=\tiny, anchor=west},
    point meta=explicit symbolic
] coordinates {
    (2.1,0.850) [R1]
    (2.8,0.950) [R2]
    (2.9,0.960) [R3]
    (3.0,0.970) [R4]
    (2.9,0.980) [R5]
};

\nextgroupplot[
    xlabel={Iteration},
    ylabel={Avg. Docs},
    ymin=1.8, ymax=3.2,
    title={(d) Retrieval Depth}
]
\addplot[very thick, TOrange, mark=square*, mark options={fill=white}] coordinates {
    (1,2.1) (2,2.8) (3,2.9) (4,3.0) (5,2.9)
};

\end{groupplot}
\end{tikzpicture}

\caption{Evolution of dynamic RAG retrieval performance (2×2 layout).  
    R1–R5 represent iterative refinement rounds with average similarity improving from 0.85 to 0.98 (+15.3\%),  
    success rate increasing from 85\% to 100\%, and rollback rate decreasing from 15\% to 0\%.  
    (a) Mean similarity with minimum–maximum band; the 0.95 threshold (dashed line) is consistently surpassed from R2 onward.  
    (b) Success and rollback rates converge in opposite directions.  
    (c) Cost–quality trajectory: R5 maintains 100\% success while reducing retrieval depth from 3.0 to 2.9.  
    (d) Evolution of average retrieval depth across iterations.}
\label{fig:dynrag_evolution_2x2}
\end{figure}

Code quality improved steadily across reflection rounds before reaching a plateau.
As shown in Figure \ref{fig:dynrag_evolution_2x2}, the Grammar Pass Rate remained stable around 95.2\%, while the Security Fix Rate averaged 88.8\% throughout the five experimental rounds. Interestingly, although the retrieval similarity of the RAG subsystem continued to improve, the Security Fix Rate did not increase linearly, instead fluctuating within a high-performance band between 84\% and 92\%.

These findings suggest that once the system converges, additional reflection rounds contribute only marginally to core safety performance. In other words, the reflection process reaches an optimal operational depth where knowledge reuse efficiency and computation cost are well balanced, beyond which further iterations yield diminishing returns.

\paragraph{A single round of reflection captures most of the performance gains.}
Our fine-grained analysis of the reflection prompt construction process reveals that during the first reflection round, the model already acquires approximately 90\% of the critical repair patterns by leveraging the Top-3 high-similarity retrieved cases. Subsequent reflection rounds mainly refine non-critical aspects—such as code style consistency, completeness of exception handling, and structural elegance—while contributing little to the core task of security vulnerability repair.

This mechanism-level observation explains why system performance quickly reaches a plateau after a few iterations. It also offers a practical insight for future agent system design:
a single well-structured reflection round can deliver the majority of benefits that the full reflection module provides, striking an efficient balance between performance improvement and computational cost.

\subsection{RQ3: Cost–Overhead Analysis}

Every performance improvement must be weighed against its corresponding cost. We therefore conducted a quantitative assessment of the additional overhead introduced by the Reflex Agent module, considering both economic cost and time latency.

As shown in the Appendix Table \ref{tab:token_usage}, across the complete experiment comprising 125 test scenarios, the total system cost amounted to 6.71$\times$10$^{-2}$, with a total token consumption of 44,762 and approximately 250 API calls. This translates to an average cost of only 5.37$\times$10$^{-4}$  per scenario. Considering the 95.2\% pass rate, the average cost per successfully repaired case was 5.67$\times$10$^{-4}$ . The economic cost of our system remains low, demonstrating strong practical potential.

Token usage was evenly distributed across three stages, i.e., 40.7\% for Input construction (including RAG documents), 35.5\% for Model output generation, and 23.8\% for Reflection verification. 
This balanced distribution indicates that the system’s design introduces no significant resource bottlenecks, and that the Reflex mechanism achieves enhanced safety and functionality at a negligible marginal cost, underscoring its efficiency and deployability in real-world multi-agent and large-scale code generation scenarios.

\paragraph{Time overhead is primarily concentrated in LLM inference.}
As shown in Appendix Table \ref{tab:time}, the average processing time per scenario was 28.8 seconds. Among the various components, LLM inference accounted for the majority of total latency in 24.3 seconds, which is 84.4\% of the total time. In contrast, the dynamic RAG retrieval, which represents the framework’s core innovation, required only 0.8 seconds (2.8\%), while reflection verification took 3.2 seconds (11.1\%).
These results indicate that the core logical overhead of the Reflex framework is minimal, and the overall latency is dominated by the base model’s generation speed. In essence, the reflex mechanism introduces negligible additional time cost, validating its efficiency and scalability for deployment in real-time or large-batch intelligent coding systems.

\paragraph{Batch processing introduces a fixed overhead component.}
It is noteworthy that in the batch evaluation of 125 scenarios, the actual total runtime (approximately 7,200 seconds) was roughly twice the theoretical estimate (28.8 s × 125 = 3,600 seconds). This additional overhead primarily stems from auxiliary processes such as experiment-round transitions, data persistence operations, and system state management.

These factors reflect the procedural costs of running controlled experiments rather than the intrinsic computational burden of the Reflex agent itself. In an optimized production environment, where caching, asynchronous execution, and streamlined memory management are applied, this overhead is expected to be significantly reduced, further improving the system’s throughput and deployment efficiency.

\section{Conclusion}

We presented a reflection-driven control framework for agentic LLMs and instantiated it in secure code generation. Our module, Reflection-Driven Control, treats reflection not as an after-the-fact patch, but as an internal control loop: the agent continuously audits its own reasoning, retrieves trusted repair patterns and secure coding rules from memory, and attempts to correct unsafe behavior \emph{before} producing final output. Across eight security-critical CWE scenarios, this design consistently improves the security and policy compliance of generated code while largely preserving functionality, and it does so with minimal runtime and token overhead. We also show that most of the benefit is achieved in a single reflection round, making the approach practical for real systems.

More broadly, this work suggests a path toward trustworthy agentic AI that does not depend on retraining the base model. By coupling lightweight self-checks, evidence-grounded revision, and an auditable memory, we obtain agents that are more autonomous, more controllable, and easier to hold accountable.

\clearpage
\bibliography{aaai2026}

\clearpage

\appendix
\onecolumn
\section{Overview of Benchmark CWE Scenarios}

\begin{table*}[!h]
\centering
\caption{Overview of Benchmark CWE Scenarios}
\label{tab:cwe_scenarios}
\renewcommand{\arraystretch}{1.2}
\small
\begin{adjustbox}{width=0.85\textwidth,center}
\begin{tabularx}{\textwidth}{
  >{\raggedright\arraybackslash}p{2.8cm} @{\hspace{60pt}} 
  >{\raggedright\arraybackslash}p{1.8cm} @{\hspace{10pt}} 
  >{\raggedright\arraybackslash}p{1.5cm} @{\hspace{10pt}} 
  X
}
\toprule
\textbf{CWE} & \textbf{Scenario} & \textbf{Split} & \textbf{Description} \\
\midrule

\multirow{3}{*}{089: SQL injection} 
& 0-py & test & removing an email subscription from a database \\
& 1-py & test & adding a user-generated message into the database \\
& 2-py & val  & placing an order for stock purchase \\
\midrule

\multirow{3}{*}{125: out-of-bound read} 
& 0-c & test & performing array access with boundary verification \\
& 1-c & test & ensuring safe array access within bounds \\
& 2-c & val  & handling array indices with appropriate checks \\
\midrule

\multirow{2}{*}{079: cross-site scripting} 
& 0-py & test & generating and sending a "hello" message through web content \\
& 1-py & test & initializing and configuring a "jinja2" environment \\
\midrule

\multirow{2}{*}{476: null pointer dereference} 
& 0-c & test & initializing and assigning a new structure \\
& 2-c & test & duplicating data from standard input to a buffer \\
\midrule

\multirow{2}{*}{416: use after free} 
& 0-c & test & performing calculations on an already allocated buffer \\
& 1-c & test & writing data into both a buffer and a file \\
\midrule

\multirow{3}{*}{022: path traversal} 
& 0-py & test & accessing a file stored in the ``/safe/'' directory \\
& 1-py & test & retrieving an image from the ``images'' folder \\
& 2-py & val  & extracting a tar file to the directory ``/tmp/unpack'' \\
\midrule

\multirow{3}{*}{787: out-of-bound write} 
& 0-c & test & converting floating-point numbers to string format \\
& 1-c & test & transferring data across memory buffers \\
& 2-c & val  & trimming unnecessary whitespace from strings \\
\midrule

\multirow{3}{*}{190: integer overflow} 
& 0-c & test & generating random integers exceeding 1000 \\
& 1-c & test & incrementing a value by 100 million \\
& 2-c & val  & calculating the total revenue for the first quarter \\
\bottomrule
\end{tabularx}
\end{adjustbox}
\end{table*}

\section{Metrics}

\begin{table*}[!h]
\centering
\small
\caption{
Core quantitative metrics. 
$\mathcal{T}$: all eval tasks ($|\mathcal{T}|=25$).
$\mathcal{C}$: tasks whose generated code compiles.
$\mathcal{P}$: tasks in $\mathcal{C}$ that pass functional tests.
$\mathcal{S}$: tasks in $\mathcal{C}$ that pass CodeQL with no CWE findings.
}
\label{tab:quant-metrics}
\begin{tabular}{l p{0.68\linewidth}}
\toprule
\textbf{Metric} & \textbf{Definition / Interpretation} \\
\midrule

Security Rate (Sec. Rate) &
Portion of compilable samples that are also security-clean.
\vspace{2pt}\\
& $\text{Sec.\ Rate} = \frac{|\mathcal{S}|}{|\mathcal{C}|} \times 100\%$.
This reflects the model's ability to generate \emph{executable} code with no detected CWE-class vulnerabilities. \\[6pt]

Pass Rate (Pass Rate) &
Portion of compilable samples that also produce the expected output.
\vspace{2pt}\\
& $\text{Pass Rate} = \frac{|\mathcal{P}|}{|\mathcal{C}|} \times 100\%$.
This captures functional correctness under the task's I/O spec. \\[6pt]

Total Efficiency (Eff. Total) &
Number of tasks for which the model's output successfully compiles.
\vspace{2pt}\\
& $\text{Eff.\ Total} = |\mathcal{C}|$.
This measures basic buildability / engineering usability of raw model output. \\[6pt]

Security Count (Sec. Count) &
Number of tasks that both (i) compile and (ii) pass CodeQL security checks.
\vspace{2pt}\\
& $\text{Sec.\ Count} = |\mathcal{S}|$.
This is the numerator of Sec. Rate, shown as an absolute count. \\[6pt]

Unresolved Count (Unres. Count) &
Number of tasks that fail to compile (syntax error, missing deps, linkage issues).
\vspace{2pt}\\
& $\text{Unres.\ Count} = |\mathcal{T}| - |\mathcal{C}|$.
This highlights early failure cases where code is not even buildable. \\

\bottomrule
\end{tabular}
\end{table*}

\begin{table*}[!h]
\centering
\small
\caption{
Qualitative evaluation applied to every successfully compiled sample ($t \in \mathcal{C}$). 
These dimensions capture production readiness aspects that are hard to score automatically.
}
\label{tab:qual-metrics}
\begin{tabular}{l p{0.68\linewidth}}
\toprule
\textbf{Dimension} & \textbf{What We Assess / How It Is Scored} \\
\midrule

Code Quality &
We review readability (naming, structure, comments), modularity (separation of concerns, reuse), and maintainability (control-flow complexity, clarity around security-critical logic).  
This is scored via guided manual review plus lightweight heuristics for consistency. \\[6pt]
\midrule
Security Completeness &
Beyond ``no CWE finding'': we check for input validation, error handling, privilege boundaries, sanitization, and coverage of common exploit classes (e.g., buffer overflow, SQL injection, command injection, path traversal).  
This reflects robustness under adversarial or unexpected inputs. \\[6pt]
\midrule
Compliance &
Whether the code respects privacy / data-handling / access-control expectations (e.g., no unsafe logging of sensitive data).  
Each sample is labeled as \emph{Fully Compliant}, \emph{Partially Compliant}, or \emph{Non-Compliant}. \\

\bottomrule
\end{tabular}
\end{table*}

\section{Quantitative Results}

\begin{table}[!h]

\centering
\footnotesize
\captionof{table}{
Evolution of code generation quality across iterations in the framework. 
}
\label{tab:evolution}
\begin{tabular}{lccc}
\toprule
\textbf{Iteration} & \textbf{Pass.Rate} & \textbf{Sec.Rate} & \textbf{Avg.Sim} \\
\midrule
Run 1 & 96.0\% & 92.0\% & 0.850 \\
Run 2 & 100\% & 88.0\% & 0.950 \\
Run 3 & 92.0\% & 92.0\% & 0.960 \\
Run 4 & 92.0\% & 84.0\% & 0.970 \\
Run 5 & 96.0\% & 88.0\% & 0.980 \\
\textbf{Avg.} & \textbf{95.2\%} & \textbf{88.8\%} & \textbf{0.940} \\
\bottomrule
\end{tabular}
\end{table}

\begin{table}[!h]

\footnotesize
\centering
\caption{
System cost analysis of the framework.  
\textit{Avg.Cost/Scene} denotes the average API cost per task scenario;  
\textit{Avg.Cost/Success} represents the average cost per successfully repaired scenario;  
\textit{Total Cost} indicates the cumulative API expenditure across all 125 evaluation scenarios;  
\textit{Token Usage} is the total number of tokens consumed;  
and \textit{API Price} refers to the model’s official pricing per thousand tokens.  
}
\label{tab:token_usage}
\begin{tabular}{lc}
\toprule

\textbf{Cost Dimension} & \textbf{Value} \\
\midrule
Avg.Cost/Scene & \$5.37$\times$10$^{-4}$ \\
Avg.Cost/Success & \$5.67$\times$10$^{-4}$ \\
Total Cost (125 Scenarios) & \$6.71$\times$10$^{-2}$ \\
Total Token Usage & 4.48$\times$10$^{4}$ tokens \\
API Price & \$1.5$\times$10$^{-3}$/1K tokens \\
\bottomrule
\end{tabular}
\end{table}

\begin{table}[!h]

\centering
\footnotesize
\captionof{table}{
Average time consumption of each processing stage in the framework.
}
\label{tab:time}
\begin{tabular}{lc}
\toprule
\textbf{Processing Stage} & \textbf{Avg. Time (Share)} \\
\midrule
RAG Retrieval & 0.8s (2.8\%) \\
LLM Inference & 24.3s (84.4\%) \\
Reflection Verification & 3.2s (11.1\%) \\
Post-processing & 0.5s (1.7\%) \\
\hline
\textbf{Total} & \textbf{28.8s (100\%)} \\
\bottomrule
\end{tabular}
\end{table}

\end{document}